# Accelerated development of $CuSbS_2$ thin film photovoltaic device prototypes

Adam W. Welch[1,2,*], Lauryn L. Baranowski[1,2], Pawel Zawadzki[1], Clay DeHart[1], Steve Johnston[1], Stephan Lany,[1] Colin A. Wolden[2], and Andriy Zakutayev[1]

1 National Renewable Energy Laboratory, Golden CO

2 Colorado School of Mines, Golden CO

Development of alternative thin film photovoltaic technologies is an important research topic due to the potential for low-cost, large-scale fabrication of high-efficiency solar cells. Despite the large number of promising alternative absorbers and corresponding contacts, the rate of progress is limited by complications that arise during solar cell fabrication. One potential solution to this problem is the high-throughput combinatorial method, which has been extensively used for research and development of individual absorber and contact materials. Here, we demonstrate an accelerated approach to development of thin film photovoltaic device prototypes based on the novel $CuSbS_2$ absorber, using the device architecture employed for $CuIn_xGa_{(1-x)}Se_2$ (CIGS). The newly developed three-stage, self-regulated $CuSbS_2$ growth process enables the study of PV device performance trends as a function of phase purity, crystallographic orientation, layer thickness of the absorber, and numerous back contacts. This exploration results in initial $CuSbS_2$ device prototypes with ~1% conversion efficiency, currently limited by low short-circuit current due to poor collection of photoexcited electrons, and a small open-circuit voltage due to a theoretically predicted, cliff-type conduction band offset between $CuSbS_2$ and CdS. Overall, these results illustrate the potential of combinatorial methods to accelerate the development of thin film photovoltaic devices with this and other novel absorbers.

*corresponding author

## 1 Introduction

Significant efforts have been directed in recent years to develop thin film solar cells based on novel absorbers, such as $Cu_2ZnSnSe_4$ (CZTS) [1], $CsSnI_3$ [2], $Cu_2SnS_3$ [3], SnS [4], $Cu_2O$ [5], $Cu_3N$ [6], $ZnSnN_2$[7], $Sb_2S_3$ [8]. Typically, the absorber studies focus on the optical properties (i.e. band gap, absorption coefficient), followed by limited research on electrical transport (carrier concentration, mobility). High-throughput theoretical [9] and -experimental [10] methods have been shown useful for accelerating such materials research in both absorbers [11] and contacts [12]. However, favorable bulk absorber properties and suitable band alignment are necessary but not sufficient for obtaining high PV device efficiencies. Many practical challenges arise from materials integration into initial photovoltaic device prototypes, including effects of materials structure (morphology, orientation) on the absorber performance, chemical reaction or interdiffusion with the underlying contacts during the absorber growth, and engineering of pinhole-free absorbers that do not delaminate from the substrate. Recently, high-throughput experimental methods have applied to optimization of 10-20% efficient CIGS with well-established PV device structure [13], as well as very novel all-oxide solar cells [14]. Combining the strengths of these two high-throughput approaches, in this paper we demonstrate combinatorial methods for accelerated development of initial PV device prototypes with novel absorbers in the previously established device structures.

$CuSbS_2$ is the novel ternary copper chalcogenide absorber that is focus of this paper. With a band gap of 1.5 eV and moderate hole doping ($10^{15}$ -$10^{18}$ $cm^{-3}$) [15-20], the $CuSbS_2$ basic electro-optical properties are similar to CIGS. However, $CuSbS_2$ has higher absorption coefficient ($\alpha > 10^5$ $cm^{-1}$ only 0.3 eV above the absorption onset) and larger effective masses ($m^*_e = 2.9$ $m_e$, $m^*_h = 3.7$ $m_e$ calculated in this work), both of which result from larger density of states due to the anisotropic $CuSbS_2$ crystal structure. The layered $CuSbS_2$ crystal structure stems from the low-valent $Sb^{III+}$ ions (group-V element in III+ oxidation sate) that adopt trigonal pyramidal coordination with a lone pair of non-bonding electrons at the apex. The $CuSbS_2$-based PV devices reported to date [18,21,22] show efficiencies of up to 3.1% [22], with the absorbers produced by sulfurization of electrode-deposited metallic stacks. In this work the $CuSbS_2$ absorbers were synthesized by RF sputtering, which is commonly used for the ZnO:Al TCO front contact fabrication in commercial CIGS devices. In the long term, this deposition strategy may lead to in-line sputtering of the entire solar cell stack on steel substrates, which together with lower demand and greater supply of Sb [23] may result in potential cost savings of $CuSbS_2$ compared to CIGS. Furthermore, at the earlier stages of the $CuSbS_2$ PV technology development (including this paper) significant cost savings result from leveraging the existing and extensive knowledge base in CIGS absorber growth (three-stage co-deposition approach [24], two-step selenization process [25]) and device fabrication (Mo back contacts, CdS/AZO front contact).

Here, we demonstrate initial prototypes of $CuSbS_2$ thin film photovoltaics (PV) using accelerated PV device development methods. We start by developing a three-stage self-regulated absorber growth process akin to CIGS, in order to control the reproducibility of the resulting PV devices. Then, we demonstrate the combinatorial approach to solar cell development by quickly exploring the PV device performance as a function of phase purity, crystallographic orientation, and thickness of the novel $CuSbS_2$ absorber. We then go on to lend some insight to the discovered trends with theoretical calculations. Finally we screen a wide range of potential back contacts for $CuSbS_2$ devices, finding Mo, including a thin $MoO_x$ charge-selective layer for thin absorber layers, to be better compared to other metals and resulting in ~1% initial PV device efficiencies. Overall, these results demonstrate an example of accelerated approach to development of initial thin film photovoltaic device prototypes from novel absorbers that would be applicable to other materials.

## 2 Methods

The $CuSbS_2$ absorber growth is performed in a vacuum chamber with $10^{-7}$ Torr base pressure, under a flow of Ar gas (3mTorr, 99.99% purity). The three 2" sputter sources loaded with one $Cu_2S$ (99.99% purity) and two $Sb_2S_3$ (99.99% purity) targets, are sputtered at 40W of RF power, resulting in absorber deposition rates of 5-10 nm/min. The final absorber thickness (0.5 – 3.0 um) was controlled by the time of the deposition (1 - 8 hours). The absorbers are deposited on heated (15° C/min) 50x50 mm stationary (not rotated) metal-coated soda lime glass (SLG) or Corning Eagle-XG glass (EXG) substrates. Prior to the absorber deposition, the back contact metals were deposited on the substrates using evaporation for Au, Pt, Pd, W, Ni, and using DC sputtering for Mo. The FTO/SLG TEC15 substrates were purchased. The $MoO_x$ charge-selective contacts were grown on a Mo electrode by 5 or 15 second dip in 30% reagent grade aqueous $H_2O_2$, and rinsed in deionized water. After the absorber deposition, the front contacts are prepared by chemical bath deposition (CBD) of CdS [26], RF sputtering of intrinsic/conductive i-ZnO/ZnO:Al (AZO) stack, and e-beam evaporation of Al metal through a shadow mask. Finally, individual device isolation is done by gentle razor blade scribing through most of the device stack all the way down to the back contact, resulting in the combinatorial PV device library with different front contacts but one common back contact.

A schematic representation of the combinatorial PV device library is shown in Fig. 1a. During the absorber deposition step of the PV device fabrication, the flux from the two $Sb_2S_3$ targets is perpendicular to the flux from the one $Cu_2S$ target. This deposition geometry results in orthogonal combinatorial gradients of the $CuSbS_2$ thickness (parallel to the $Cu_2S$ flux) and crystallographic orientation or phase purity (parallel to the $Sb_2S_3$ flux). The rest of the PV device layers, including metal back contact and CdS/TCO/metal front contact are spatially uniform across the combinatorial PV device library. A typical cross-section of one device from the library is shown in Fig. 1b, where with these different layers can be seen.

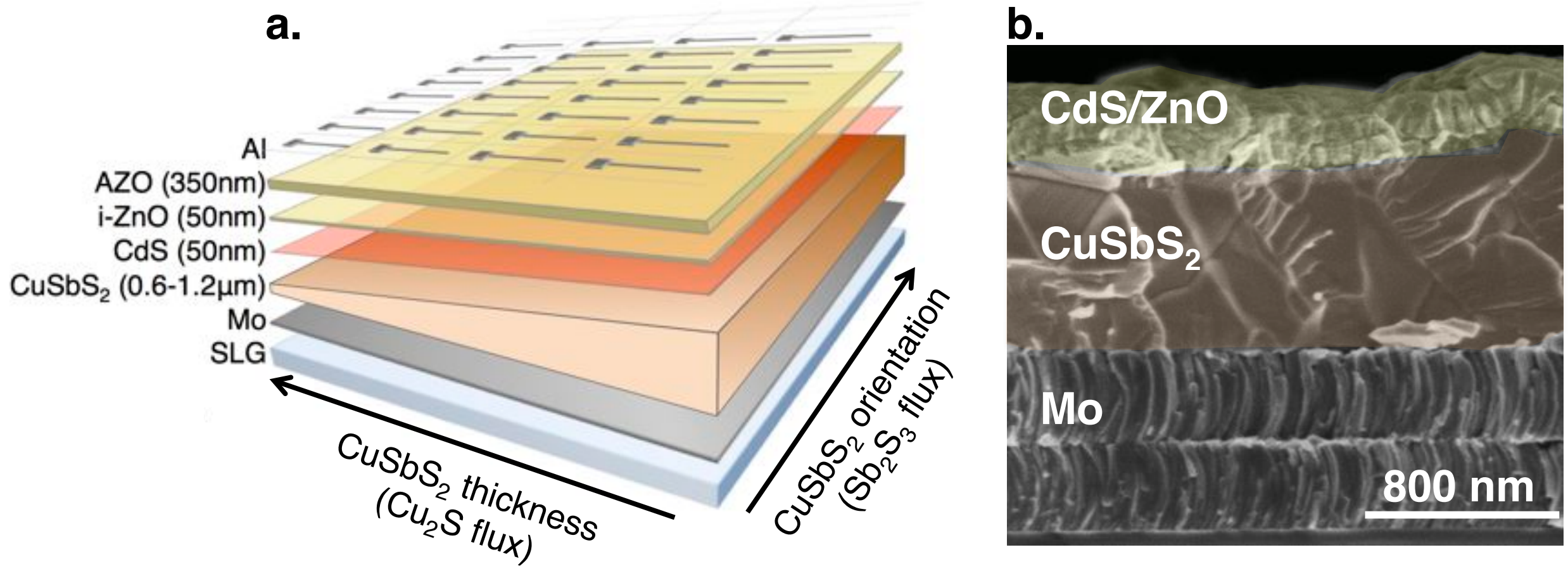

**Figure 1:** (a) A schematic diagram of a combinatorial PV device library, showing the orthogonal gradients in the $CuSbS_2$ absorber thickness and its crystallographic orientation, as well as the 4x11 grid of Al front contacts. (b) A false color cross section scanning electron micrograph of a typical device, showing the ZnO/CdS front contact, $CuSbS_2$ absorber, and Mo back contact layers.

The shadow mask used in this study has 4 rows and 11 columns of contacts, resulting in 44 individual devices on the PV device library. As shown in Fig. 1a, this 4x11 shadow mask can be placed such that there are 11 devices along the $CuSbS_2$ crystallographic orientation or phase purity gradients, to quickly study the effects of these absorber properties on the device performance. Alternatively, in the combinatorial PV device libraries with no orientation/purity gradients, the 11 devices can be placed parallel to the $CuSbS_2$ thickness gradient, in order to efficiently optimize the thickness of the absorber. Finally, if the 11 devices are placed perpendicular to the thickness gradient for the libraries with no orientation/purity gradients (in other words, along the nominally uniform direction of the sample), the results can be used for the statistical analysis, with 11 nominally equivalent devices per set.

The performance of combinatorial PV device libraries is measured using a custom, automated and spatially resolved, current-voltage (J-V) characterization under simulated AM1.5G illumination on a water-cooled stage (25° C). The resulting J-V data and the device parameters ($J_{sc}$, $V_{oc}$, FF, $\eta$, $R_{sh}$, $R_s$) are loaded, processed and plotted using customized algorithms implemented in the Igor-Pro software package. The measurements of absorber phase purity and crystallographic orientation are performed using spatially resolved x-ray diffraction (XRD, Bruker D8 Discover). The absorber composition and thickness are mapped using x-ray fluorescence (XRF, Fischerscope XDV-SDD). All these combinatorial measurements are performed on the same 4x11 grid of points as used for device fabrication, in order to correlate the device performance with the specific materials properties.

For the selected devices, we also performed manual single-point characterization of the devices and the absorber material to get better scientific understanding of the performance-limiting factors. The single-point capacitance-voltage (C-V) measurements were performed on a custom setup with LabView software. The external quantum efficiency (EQE) was measured using an Oriel IQE 200 instrument. The morphology of the samples was studied using scanning electron microscopy (SEM) on a JEOL JSM-7000F SEM. The devices were also characterized by dark lock-in thermography (DLIT) to investigate the origin of variations in their shunt resistance. The absorption coefficients were measured using diffuse reflectance and transmittance measurements (Cary 5000i) of the samples deposited on EXG glass substrates.

Density functional calculations of the $CuSbS_2$ surface energies, density of states and ionization potentials were performed with Perdew-Burke-Ernzerhof [27] exchange correlation functional using VASP code [28]. An effective on-site potential $U = 5$ eV was applied to Cu *d*-states [29]. The ionization potentials and electron affinities from DFT were corrected using the band edges taken form GW calculations [30]. In order to compensate for the overestimated d-orbital energies in GW, an on-site d-state potential of $V_d = -2.8$ eV is applied to the Cu *d*-states [31]. The $CuSbS_2$ effective masses were estimated from the calculated DOS using energy smearing that corresponds to 1000K effective temperature.

## 3 Results and Discussion

### *3.1 Three-stage self-regulated growth process*

We have recently demonstrated one-stage synthesis of phase pure $CuSbS_2$ directly onto heated glass substrates by self-regulated growth approach [15]. We found that there is a 75° C processing window, where excess flux of $Sb_2S_3$ sublimes from the growing phase-pure $CuSbS_2$ film, thereby avoiding $Sb_2S_3$ precipitates (lower T) or decomposition to the $Cu_{12}Sb_4S_{13}$ phase (higher T). Within this window of phase pure $CuSbS_2$ growth, the hole concentration can be controlled by adjusting the $Sb_2S_3$ over-flux and substrate temperature. For the $Sb_2S_3$ flux that we used, the smallest hole doping found at the lowest temperature (350° C) was $10^{15}$ $cm^{-3}$, and it increased with increased substrate temperature up to $10^{18}$ $cm^{-3}$ (425° C). However, at these high temperatures, it became difficult to reproducibly control the $CuSbS_2$ growth process due to proximity to the $Cu_{12}Sb_4S_{13}$ nucleation and decomposition. In addition, upon transferring the self-regulated growth process to metallic back contact that is required for device fabrication, we found that the effective temperature at the surface of the substrate is ~25° C higher than that for the glass, probably due to differences in thermal properties of the glass substrate and the metal film.

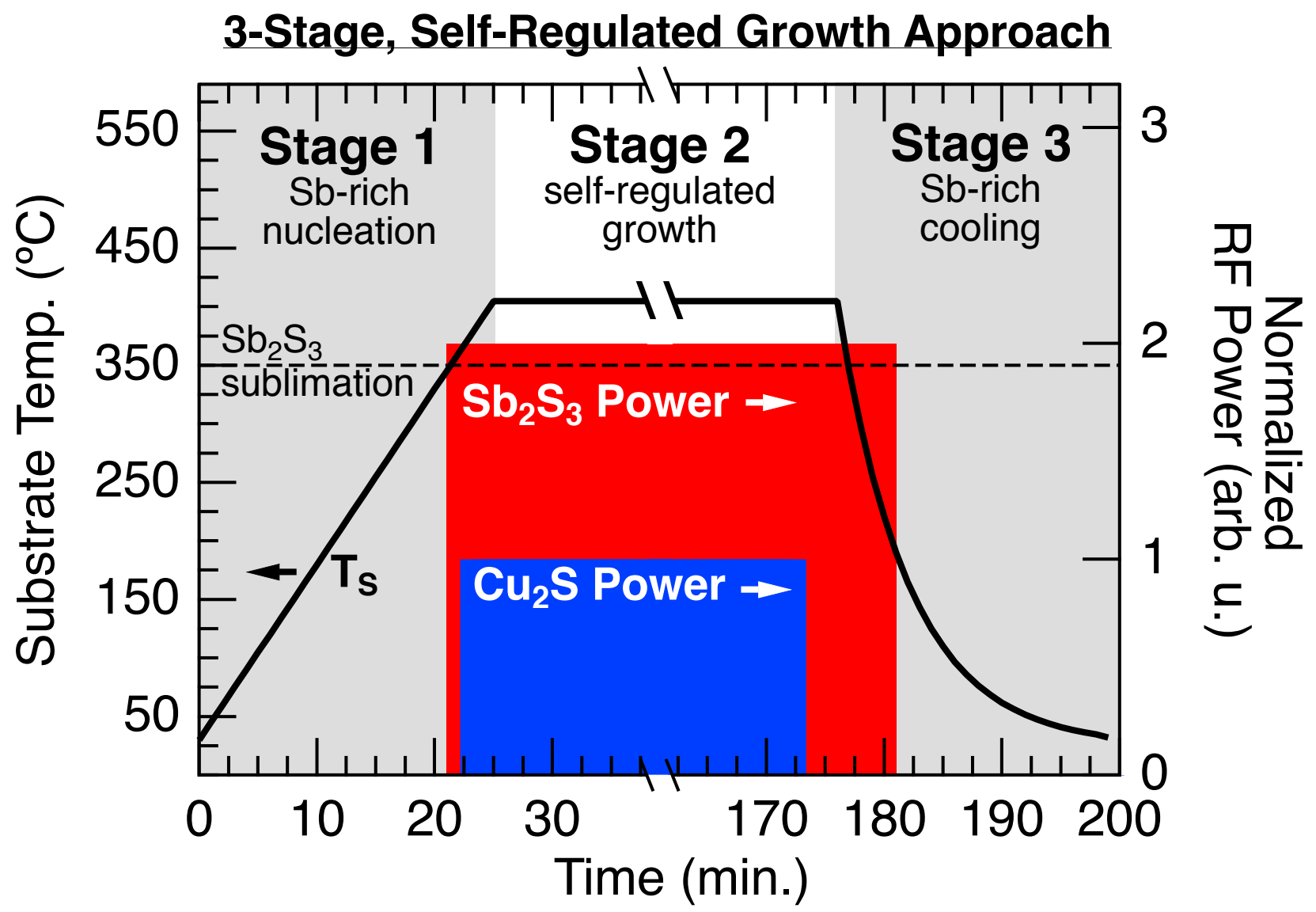

**Figure 2:** Three-stage self-regulated growth process that enables $CuSbS_2$ synthesis control at higher substrate temperatures. The first and the third stage are Sb-rich, and the second stage is performed in the self-regulated growth mode.

To enable the controlled reproducible synthesis of $CuSbS_2$ at higher substrate temperatures, we developed a three-stage self-regulated growth process (Fig. 2). It ensures that the $CuSbS_2$ film is maintained under the $Sb_2S_3$-rich conditions to avoid decomposition to detrimental $Cu_{12}Sb_4S_{13}$ at the higher synthesis temperatures.

*Stage-1:* During the first Sb-rich stage, both $Cu_2S$ and $Sb_2S_3$ sources are opened while the substrate is still heating. The lower initial temperature produces a Sb-rich $CuSbS_2$ seed layer with some $Sb_2S_3$ precipitates, ensuring against nucleation of the Cu-rich $Cu_{12}Sb_4S_{13}$ phase. As the temperature increases, these $Sb_2S_3$ precipitates sublime or react with $Cu_2S$ to form $CuSbS_2$. Fixing the substrate temperature and achieving Sb-rich nucleation by starting the $Sb_2S_3$ flux slightly before the $Cu_2S$ flux did not lead to similar phase pure films.

*Stage-2:* Once the deposition temperature is reached, the $Sb_2S_3$ precipitates sublime out of the seed layer, presumably allowing the $CuSbS_2$ grains to fill in the leftover voids, as will be discussed in more detail below. The remainder of this main stage is simply the completion of film growth to the desired thickness, while maintaining the $Sb_2S_3$ rich atmosphere to achieve the self-regulated growth regime described in [15].

*Stage-3:* Finally, in the third Sb-rich stage, the $Cu_2S$ source is shut off and the $Sb_2S_3$ sources remain open while the substrate cools to just above the $Sb_2S_3$ sublimation temperature. This ensures a high chemical potential of $Sb_2S_3$ during the short period after turning off the heat source, but before substrate temperatures are low enough to avoid decomposition.

Note that our three-stage self-regulated growth process has similarities and difference with the widely-used CIGS three-stage synthesis approach [32]. On one hand, our process starts and ends with the Cu-poor stages, just like the CIGS process. On the other hand, the main stage of our process is performed under self-regulated growth conditions in excess of $Sb_2S_3$ vapor, whereas the second stage of the CIGS process is Cu-rich. This difference results from the need to avoid formation of the Cu-rich neighboring $Cu_{12}Sb_4S_{13}$ phase in $CuSbS_2$, which is very difficult to convert back to $CuSbS_2$, probably due to large asymmetric kinetic barriers that separate the tetrahedrite ($Cu_{12}Sb_4S_{13}$) and chalcostibite ($CuSbS_2$) structures. In contrast, in the case of CIGS, the $Cu_2Se$-to-CIGS transformation is topotactic and hence facile, enabling the Cu-rich synthesis and the resulting grain growth during the second stage of the process, followed by conversion back to CIGS during the third stage.

The JV and EQE analysis of the PV devices prepared with the three-stage self-regulated $CuSbS_2$ absorber growth process are shown in Fig. 3, as a function of the second stage growth temperature. For each of these experiments, we used our combinatorial approach to fabricate sets of 11 nominally uniform PV devices with 1.5 micron thick absorber layers, (mask orientation as shown in Fig. 1a, but without the crystallographic orientation gradient), in order to access the statistical significance of the performance improvements (Table I). The JV curve of the sample deposited at 325° C displays some rollover (Fig. 3a), perhaps due to either poor collection at the back contact, or the presence of $Sb_2S_3$ impurities that had not sublimed after the first lower temperature stage (Fig. 3a inset). For the higher growth temperatures (350-400◦ C), $V_{oc}$ remains constant at about 300mV, but the $J_{sc}$, FF, and efficiency rise with increasing T due to reduced series resistance (Table I). This result is consistent with our previous work [15], showing higher carrier concentrations for higher growth temperatures. Also, note that the increase in $J_{sc}$ and FF are correlated due to the low overall photoresponse: any increase in photocurrent leads to diode turn-on voltage moving deeper into the fourth quadrant of the JV graph, producing a higher fill factor.

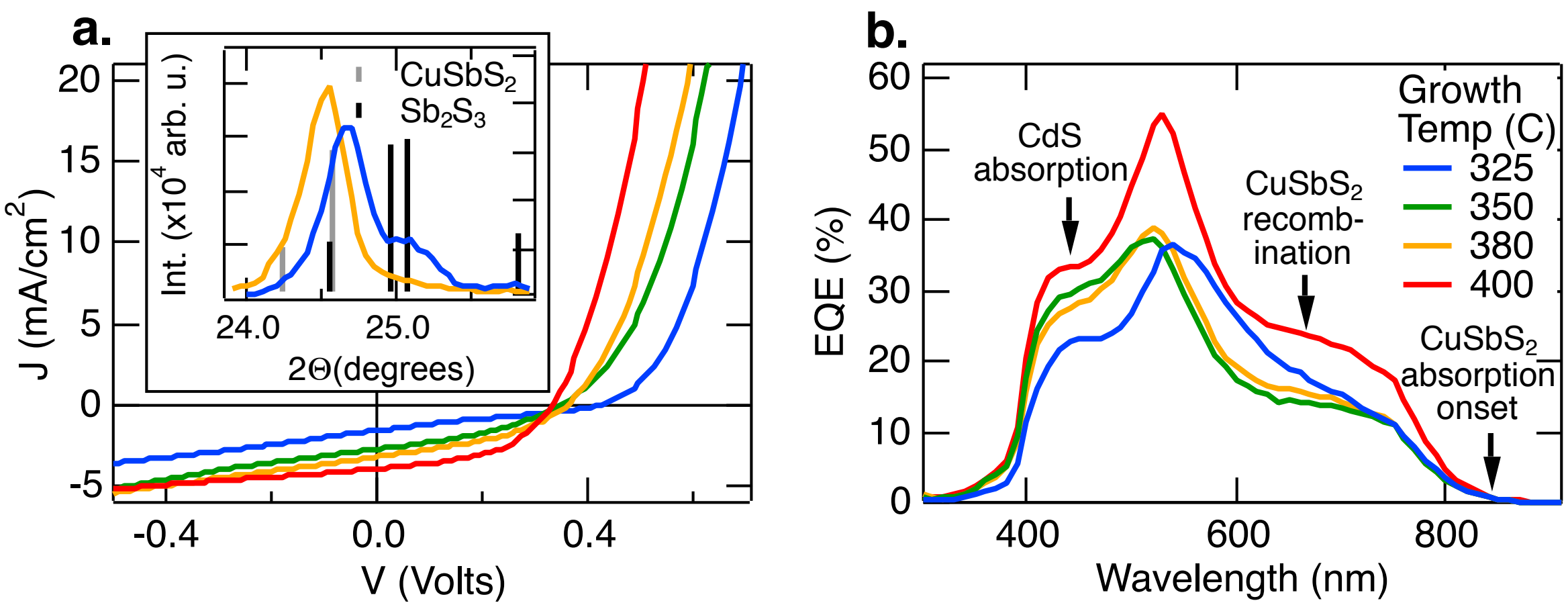

**Figure 3:** (a) JV and (b) EQE response of the $CuSbS_2$ devices as a function of the second stage growth temperature (Fig. 2). The inset in (a) shows XRD peaks associated with $Sb_2S_3$ precipitates for growth below 350°C.

Despite the statistically significant increase in the performance with increasing growth temperature (Table I and Figure 3), all samples display very low photocurrent in the 2-4 mA/cm$^2$ range. The EQE magnitude is consistent with these JV results (2-4 mA/cm$^2$ integrated current). The EQE shape (Fig. 3b) suggests poor minority carrier collection from the bulk of the absorber for all the studied growth temperatures. Specifically, wavelengths which penetrate deeper into the film, (600-700 nm), show a sharp decline in charge carrier collection, with EQE in the 10-30% range. There is also an expected 25-35% EQE notch in the 400-500 nm wavelength range associated with parasitic absorption by the CdS buffer layer. Between these collection- and absorption-limited spectral regions, EQE reaches a maximum of 33-55% at ~550 nm. Finally, in the 720-820 nm range the EQE is limited by insufficient carrier generation in the $CuSbS_2$ absorber, consistent with the ~1.5 eV optical bandgap of this material.

**Table I:** The average JV parameter and their standard deviations for 11 devices, (excluding few shunted cases), as a function of the second stage growth temperature.

| $T_{substrate}$ (C) | $J_{SC}$ (mA/cm$^2$) | $V_{OC}$ (mV) | FF (%) | η (%) | $R_S$ (Ω cm$^2$) | $R_{Sh}$ (Ω cm$^2$) |
|---|---|---|---|---|---|---|
| 325 | 2.14±0.36 | 390±36 | 25.9±1.3 | 0.21±0.04 | 263±50 | 417±90.5 |
| 350 | 2.76±0.18 | 284±52 | 31.2±2.4 | 0.25±0.07 | 154±14 | 381±97.8 |
| 380 | 3.28±0.76 | 295±41 | 34.4±4.8 | 0.33±0.09 | 103±23 | 312±108 |
| 400 | 3.87±0.64 | 303±39 | 50.2±2.3 | 0.58±0.06 | 40±7.4 | 336±217 |

Comparison of these EQE results with the absorption coefficient from optical measurements and the depletion width from C-V measurements (see supplementary Fig. S1) suggests that even the carriers generated within the space charge region do not get collected with unity probability. The CV experiments suggest a depletion width on the order of 100nm inside of the $CuSbS_2$ PV device. A simple Beer-Lambert analysis indicates that 64% of the 600 nm photons will be absorbed in these first 100nm of $CuSbS_2$, given the $10^5$ cm$^{-1}$ absorption coefficient at this wavelength [15]. This 64% carrier generation number strongly contrasts with the measured 20% EQE at this wavelength (Fig. 3b), suggesting that carrier recombination occurs even within the space charge region, where their collection is enhanced by drift due to built-in electric field. These observations call for further improvements in the $CuSbS_2$ absorber quality, in particular, a reduction in bulk defect concentration.

### *3.2 Combinatorial studies of PV absorber material properties*

Different PV absorber material parameteres, such as layer thickness, crystallographic orientation (and resulting morphology), and phase purity (depending on stoichiometry), are important parameter that affects the performance of the PV devices. For some PV technologies like CIGS [33], CZTS [34], or $Cu_2SnS_3$ (CTS) [35] significant deviations from the nominal absorber stoichiometry can be beneficial (e.g. Cu-poor absorbers) or detrimental (e.g. Cu-rich absorbers). For CdTe or GaAs, no such deviations are possible without sacrificing the absorber phase purity. In this sense, $CuSbS_2$ is more similar to CdTe and GaAs rather than CIGS, CZTS, CTS and other Cu-based chalcogendies, since $CuSbS_2$ is a line compound that cannot tolerate large deviations in stoichiometry [15]. Another feature that sets $CuSbS_2$ aside from traditional tetrahedrally-bonded thin film absorbers is its layered crystal structure. This anisotropy makes the crystallographic orientation of the $CuSbS_2$ absorbers (and the resulting differences in film morphology) an important parameter that may affect the resulting PV device performance. The anisotropic crystal structure of $CuSbS_2$ absorber also leads to the high absorption coefficient, but at the expense of larger electron- and hole effective masses. This may lead to a different absorption/collection trade-off optimum than is found for CIGS, suggesting that the $CuSbS_2$ layer thickness is another important engineering parameter that needs to be optimized, along with the phase purity and the crystallographic orientation.

*3.2.1 Phase purity and composition*

To study the effects of phase purity of the $CuSbS_2$ absorbers (controlled by composition) on the PV device properties in combinatorial way, we placed the 11 devices parallel to the gradients in phase purity; a similar geometry has been used for the combinatorial studies of the $CuSbS_2$ crystallographic orientation on the PV device performance, as shown in Fig. 1a. The combinatorial gradient in phase purity was achieved by eliminating the third, Sb-rich stage of the three-stage self-regulated absorber growth process (Fig. 2), and hence letting the $CuSbS_2$ absorber partially decompose into detrimental $Cu_{12}Sb_3S_{14}$ upon cooling. We note that both phase-purity and crystallographic orientation gradients were quite difficult to control, but nevertheless, when present, they provided a quick way to study the effects of these important parameters on the PV device performance.

Figure 4 shows the effect of the Cu-rich impurity phase ($Cu_{12}Sb_4S_{13}$), controlled by cation stoichiometry of the absorber layer, on the PV device performance. Both $V_{oc}$ and $J_{sc}$ decline rapidly with increasing Cu content in the films (Fig. 4a), which correlates with the presence of $Cu_{12}Sb_4S_{13}$ inclusions (Fig. 4b). When a substantial amount of the $Cu_{12}Sb_4S_{13}$ impurities are present, the devices shows no photoresponse due to shunting that results from the high conductivity of $Cu_{12}Sb_4S_{13}$ [15]. Quantification of this performance/purity correlation (Fig. 4c) suggests that small level of $Cu_{12}Sb_4S_{13}$ impurities are tolerable in the $CuSbS_2$ absorber (incomplete shunting throughout the absorber thickness), but the performance for such devices is quite poor. In a similar but less severe way, the Sb-rich impurity phase also results in deterioration of the PV device performance (Fig. 3a). Together, these observations reinforce the importance of development of the three-stage self-regulated growth process (Fig. 2) that leads to stoichiometric phase-pure $CuSbS_2$ thin films, in contrast to CIGS, CZTS or CTS where large deviations in stoichiometry are allowed and potentially beneficial to device performance.

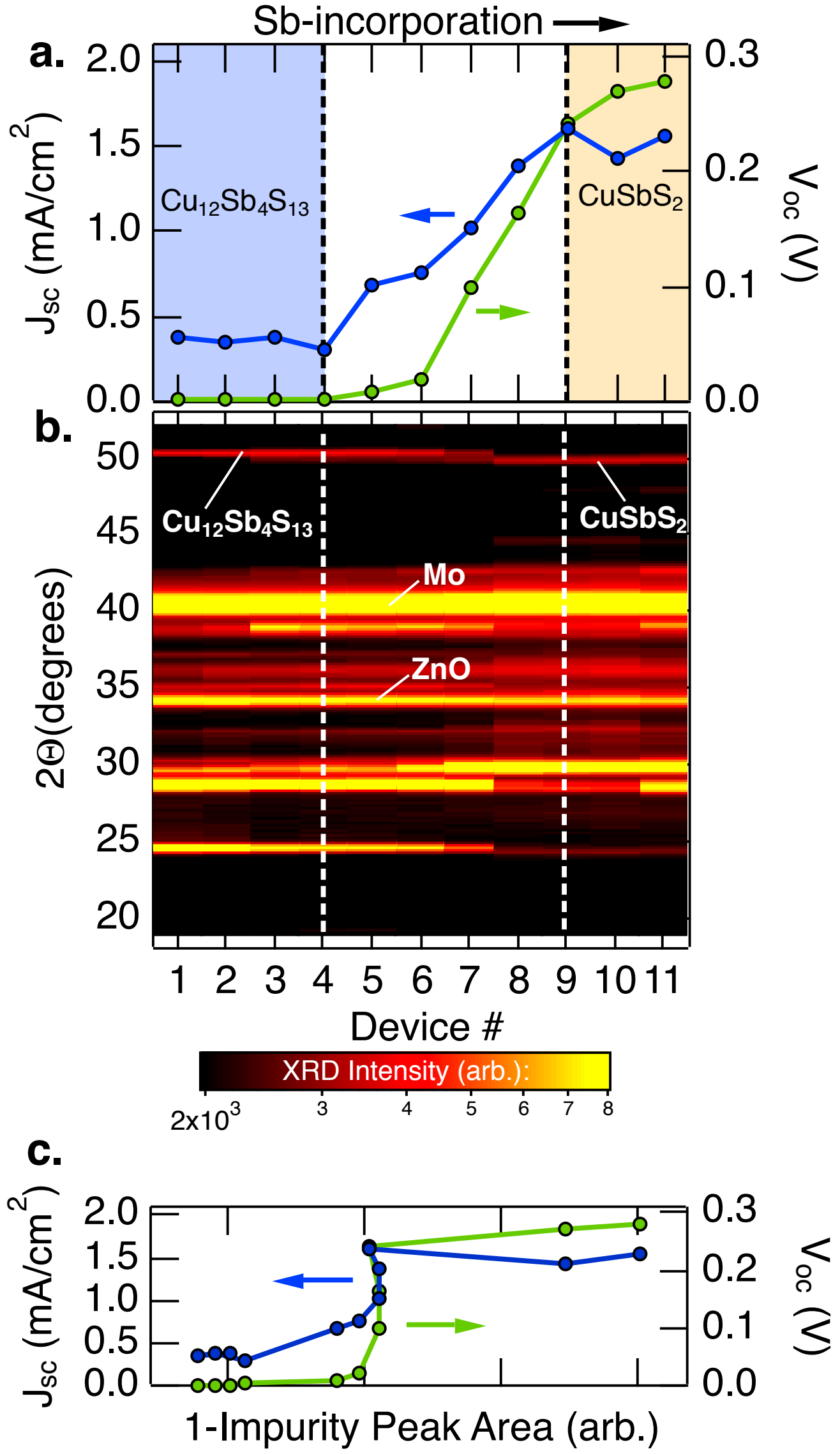

**Figure 4:** (a) $J_{sc}$ & $V_{oc}$, and (b) XRD phase purity maps of $CuSbS_2$ PV devices as a function of absorber composition. (c) Quantification of the PV performance vs. phase purity trends by integration of the $Cu_{12}Sb_4S_{13}$ impurity XRD peak area, (x-axis is 1-impurity peak area to match Sb-incorporation trend in a & b).

### *3.2.2 Crystallographic orientation and morphology*

Adjusting the timing and temperature of the first stage of the three-stage self-regulated growth process can control the crystallographic orientation and film morphology. During the first stage, as the substrate temperature rises, the $Sb_2S_3$-rich impurities will sublime; presumably leaving behind voids between $CuSbS_2$ grains with different orientations. We hypothesize that the [001]-oriented $CuSbS_2$ grains are able to grow quicker in directions parallel to the substrate plane to infill these voids, since the orthogonal [100] and [010] $CuSbS_2$ surfaces have more dangling bonds that can readily attach the incoming atoms. This overgrowth process results in large [001] oriented dendritic grains shown in Fig. 5a (left). However, if the first stage starts closer to the $Sb_2S_3$ sublimation temperature, very few voids are present, resulting in randomly oriented grains and rougher surface morphology, as shown in Fig. 5a (right). The balance between these two possible processes strongly depends on the small difference in temperature and timing of the first stage, which can be slightly different across the substrate, leading to the combinatorial gradient in the degree of crystallographic orientation and the film morphology. The effect of the resulting crystallographic orientation on device performance can be studied when gradients in orientation are present.

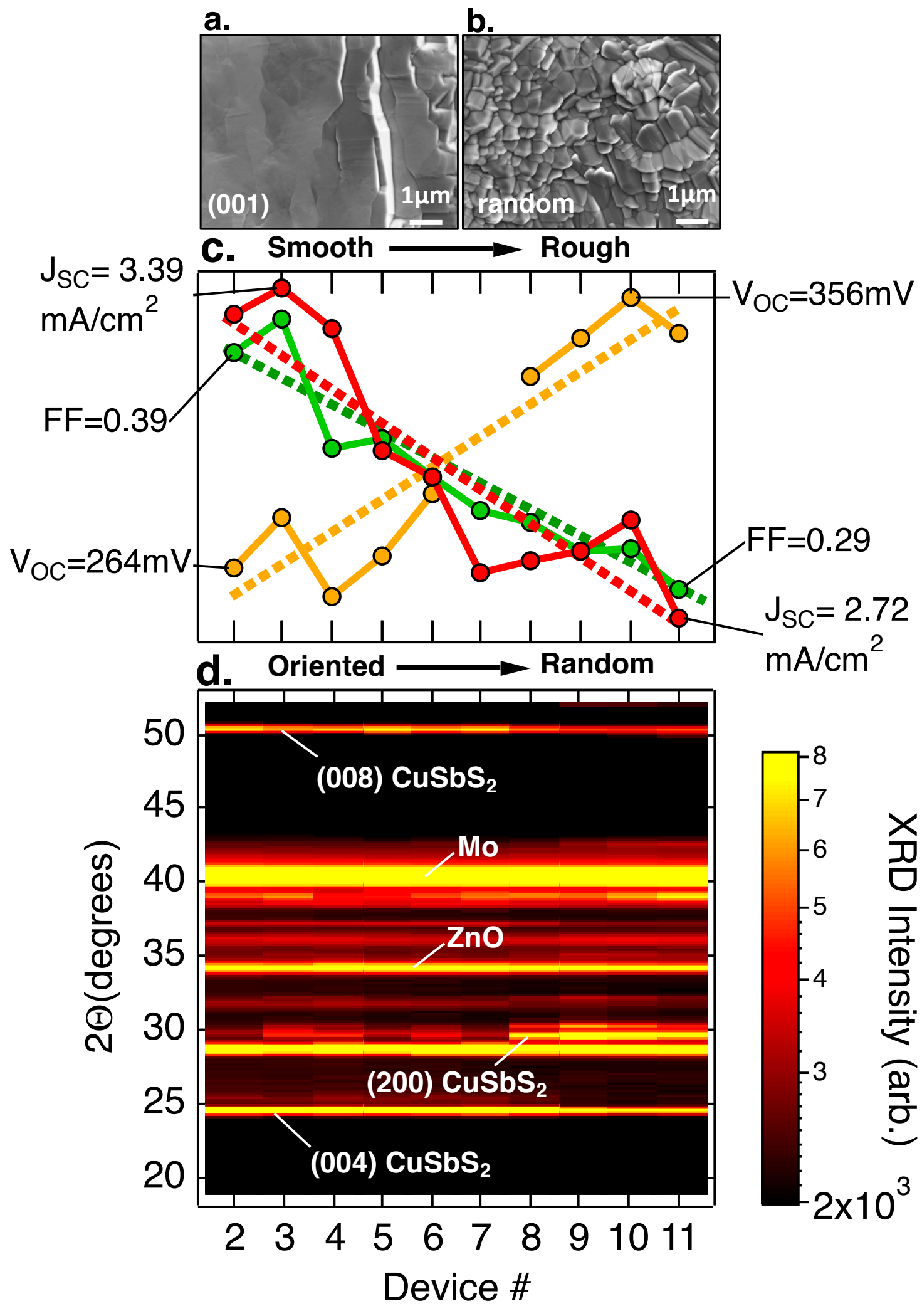

**Figure 5:** SEM images of (a) [001] oriented and (b) randomly oriented thin films. PV device performance as a function of $CuSbS_2$ crystallographic orientation and film morphology, showing how $J_{sc}$ & FF increase, and $V_{oc}$ declines with increasing [00L] orientation. The dashed lines are guides to the eye.

As shown in Fig. 5, crystallographic orientation (and resulting morphology) of $CuSbS_2$ absorbers grown at 350-380°C has strong effects on both $V_{oc}$ and $J_{sc}$ & FF, but these effects are opposing each other such that the overall device efficiency remains approximately the same. As the crystallographic orientation transitions from random to 00L-oriented, and the resulting morphology changes from relatively quite rough to smooth, the $J_{sc}$ & FF increase and the $V_{oc}$ decreases. Both of these trends can be rationalized by the change in the electron affinities and electronic states of the different $CuSbS_2$ surfaces calculated from first-principles, as discussed below.

As shown in Fig. 6a, surface calculations indicate that $CuSbS_2$ has two low energy terminations: the (001) and (010) with surface energies of 12.4 and 14.6 meV/Å respectively, possibly explaining the [001] orientation of the deposited films in Fig. 5; the (100) plane (23.1 meV/Å) as well as the (101) and (011) planes (~30 meV/Å) have higher surface energies. The electronic structure of the lowest-energy (001) and (010) surfaces indicate presence of surface states: the (010) shows 2-3 times higher density of surface states than the (001) surface, consistent with chemically benign lone-pair termination of (001). Figure 6b shows the calculated ionization potentials and electron affinities for the $CuSbS_2$ and compares them with values for CdS buffer, suggesting cliff-type band alignment with CdS in the 0.8 – 1.4 eV range depending on the $CuSbS_2$ surface orientation. The large difference in $CuSbS_2$ electron affinities depending on the surface orientation and the relatively high position of its CBM compared to CdS are both notable but not unexpected, since such effects have been reported before for other materials with lone pairs, such as SnS [36].

To summarize one possible theoretical explanation (Fig. 6) of the experimental results (Fig. 5), the increase in $J_{sc}$ (and hence FF) with increasing crystallographic orientation could be due to a decrease in number of interface states between the CdS buffer layer and larger-grain $CuSbS_2$ [001] surfaces rich with non-bonding Sb lone pairs. The decrease in $V_{oc}$ with increasing crystallographic orientation suggests an increasing cliff-type conduction band offset between CdS and the $CuSbS_2$ [001] surfaces, which are expected to have a higher conduction band position compared to the other surface orientations (Fig. 6a). However, we note that alternative explanations are also possible, since the $J_{sc}$ and $V_{oc}$ trends in Fig. 5 are observed for the relatively low efficiency devices; more efficient devices would be needed to conclusively support or rule out the theoretically predicted scientific effects in Fig. 6.

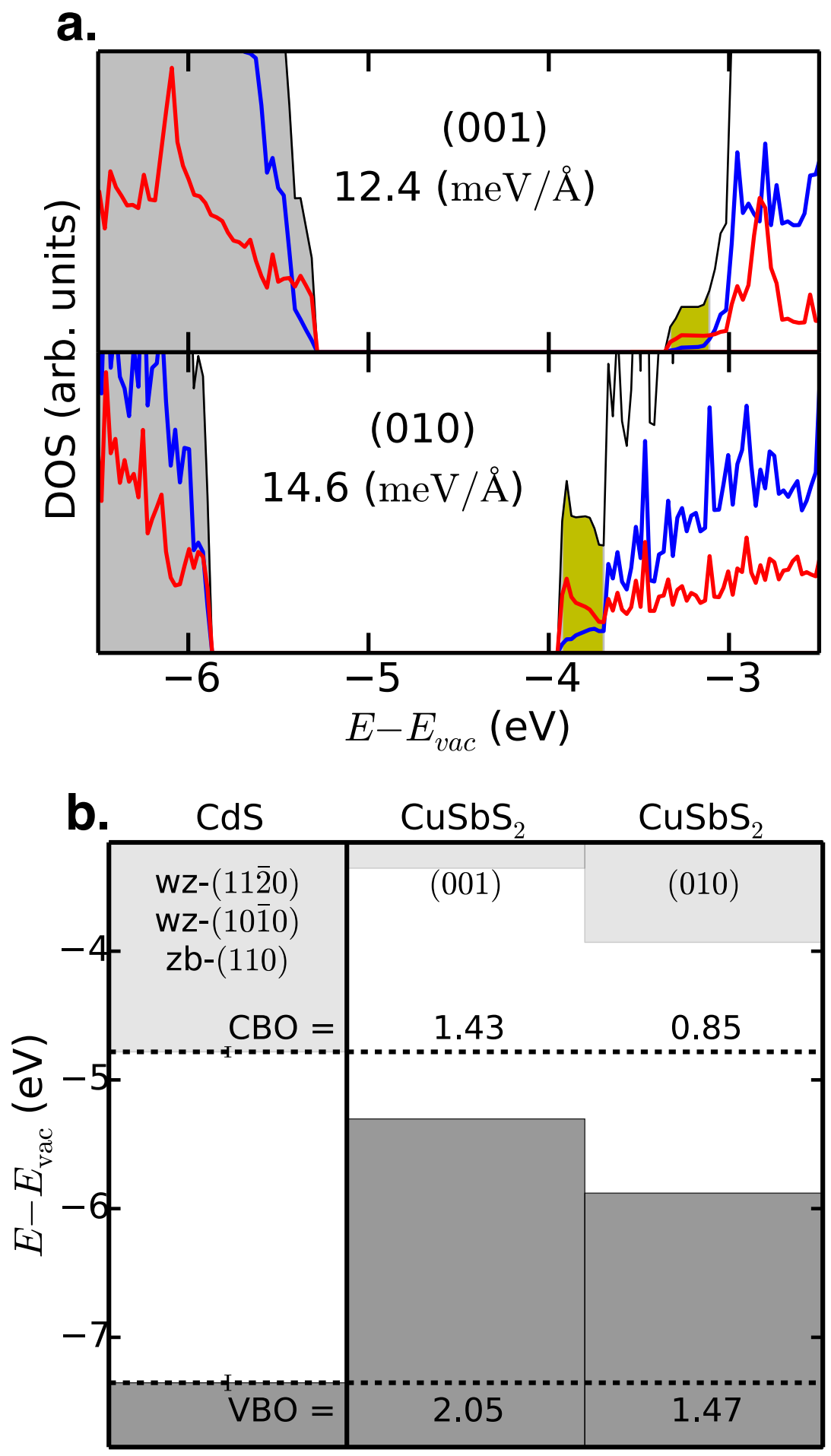

**Figure 6:** Calculated (a) density of states of $CuSbS_2$ surface slabs (normalized by surface area), and (b) band offsets for $CuSbS_2$/CdS. The [00L] planes have the most offset, resulting in reduced $V_{oc}$, while they also have the lowest energy/fewest surface states. The red and blue lines in (a) are projections onto the surface and bulk, respectively, with yellow highlighting the surface states. The CdS band positions in (b) are average over the wurzite (1120), (1010), and zinc-blende (110) orientations.

### *3.2.3 Absorber layer thickness*

Absorber thickness, controlled in second stage of the three-stage self-regulated growth process, is an important engineering parameter to be optimized for any solar cell. The thickness controls the absorption-collection trade-off: the absorber needs to be thick enough to absorb all the photons (and avoid shunts), yet thin enough to collect all the photo-generated charge carriers. Here, the thickness of the $CuSbS_2$ absorber was optimized by placing the 11 devices parallel to the thickness gradient (4x11 mask orientation orthogonal to what is shown in Fig. 1a, and without the absorber crystallographic orientation gradient). This was done on three separate combinatorial PV device libraries with different ranges of the thicknesses: 0.6 -1.0 $\mu$m thick for the first library with the absorber deposited for 2.0 hours, and 1.0 – 1.5 $\mu$m for the library grown for 4.0 hours, and a third library (without any gradients) grown for 8 hours with rotation at 20 rpm (rotation was required to ensure against gradients in orientation or phase purity which become more difficult to eliminate for longer growth periods, but also eliminated the thickness gradient for this library).

As shown if Figure 7b, the device parameters ($J_{sc}$, $V_{oc}$, FF and efficiency) show some trends with absorber thickness, is the edge effects and shunting are taken into account. In Fig. 7, devices near the edge of a library, which typically had larger variations in $J_{sc}$, are marked with open circles. Additionally, the shunt resistance, which was found to greatly affect $V_{oc}$, was used to color each point, with black points showing the highest values, and therefore the best diode behavior. Once accounting for these effects, the trends of $J_{sc}$ and $V_{oc}$ with changing absorber thickness can be identified.

As shown in Fig. 7a. the $J_{sc}$ increases from 1 to 3 mA/cm$^2$ with increasing absorber thickness in the 0.6 – 1.5 μm range, with reduced slope higher absorber thicknesses. We attribute this trend to decreased recombination in the top layer of the thick film, rather than increased absorption in the entire thickness of the film, because the studied thickness range is larger than the absorption depth of $CuSbS_2$ (90% of 600-nm photons in the first 0.2 μm of the absorber, Fig. S1). The $V_{oc}$ also increases with increasing thickness (Fig. 7b), going from ~0 V at 600nm, to 0.35 V at ~1 um, but in contrast to $J_{sc}$, this trend then flattens out, or even perhaps start to decline, at thicknesses higher than 1 μm. We attribute the initial $V_{oc}$ increase to a transition to a continuous absorber film with no pinholes. FF was analyzed but remained relatively constant in the 30% range for all thicknesses. Together, the constant FF combined with the $V_{oc}$ / $J_{sc}$ trends point to an "optimal" ~0.3% efficiency at 1.0 - 1.4 $\mu$m absorber thickness (Fig. 7c), which is thinner than that for typical for CIGS cells (~1.5-2.5 $\mu$m). In part, this can be attributed to higher absorption coefficient and larger effective masses of $CuSbS_2$ ($m^*_e = 2.9\ m_e$, $m^*_h = 3.7\ m_e$) compared to $CuInS_2$ (5-10x lower in the same theoretical approximation). Also it appears that the $CuSbS_2$ absorber thickness could be further reduced, but only if pinhole-free layers with lower defect densities at the initial stages of growth could be obtained.

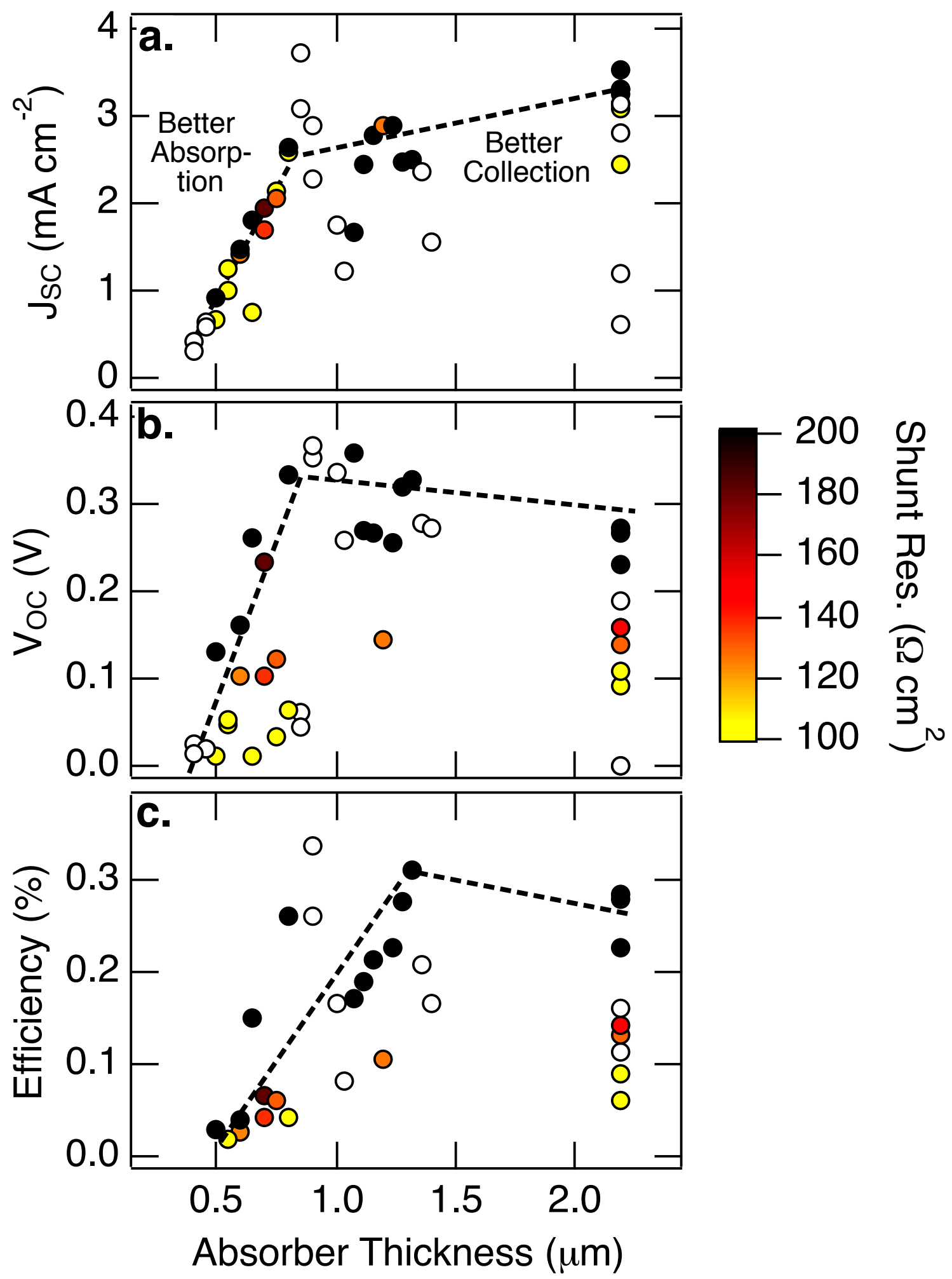

**Figure 7:** Device performance parameters, including (a) $J_{sc}$ , (b) $V_{oc}$ and (c) efficiency, as a function of $CuSbS_2$ absorber thickness. $J_{sc}$ increase with absorber thickness in the 0.6 – 1.5 $\mu$m range, but $V_{oc}$ and efficiency saturate at 1.0 $\mu$m and 1.4 $\mu$m respectively. The dashed lines are guides to the eye. Open circles are devices at the edge of a library.

### *3.3 Optimization of Back Contact*

The choice of metallic back contact is an important consideration for PV device fabrication, from both physical and chemical points of view. From the point of view of device physics, the back contact in a finished cell must efficiently collect majority charge carriers, and for some device architectures reflect the minority charge carriers. Additionally, when employing the substrate architecture, the back contact must allow for efficient nucleation, growth, and adhesion of the absorber layer. The most common back contact used for CIGS and CZTS is Mo, but other materials have also been considered [34]. Electrically, the valence bands of CZTS, CIGS and $CuSbS_2$ are deeper than the Fermi level of Mo [35], which should result in a 0.5 eV barrier to hole transport and negatively affect the PV device performance. However, it is known that deposition of CIGS and CZTS absorbers on Mo often results in formation of a thin Mo-chalcogenide layer, mitigating this problem [36]. It is likely but not guaranteed that a similar effect would occur for the $CuSbS_2$ absorber considered in this paper, so a more rigorous selection of the back contacts is needed. Here, for each of the studied back contacts we use the combinatorial approach to fabricate 11 nominally identical devices with 1.2 $\mu$m absorber thickness, allowing for statistical checks of the results.

Table II summarizes the results of PV devices with the same $CuSbS_2$ absorber but different metallic back contacts. PV devices with photoresponse were obtained only on Pt or Mo back contacts. The Mo electrode provided better current collection ($J_{sc}$), quasi-Fermi level splitting ($V_{oc}$), and diode quality (FF), despite deeper work function of Pt that should lead to better majority charge carrier collection. However, we observed that $CuSbS_2$ failed to adhere to Pt at higher substrate temperatures (>350° C), so it is possible that a less intimate electric contact between Pt and $CuSbS_2$ exists even at moderate substrate temperatures (350° C), leading to poor majority charge carrier transport. In addition, we noticed that the $CuSbS_2$ PV device on Pt had lower shunt resistance compared to the devices on Mo (Table II). Using dark lock-in thermography (DLIT), we determined that the lower shunt resistance was do to the isolation method, where the softer back contact Pt metal was displaced by the scribe putting it in touch with the front-contact TCO layer (see supplementary Fig S2).

**Table II:** The average PV device efficiency parameters and standard deviations (excluding completely shunted device) for different back contacts.

| Back Contact | $J_{sc}$ (mA/cm$^2$) | Voc (mV) | FF (%) | η (%) | $R_{sh}$ (Ω cm$^2$) |
|---|---|---|---|---|---|
| Mo | 3.53±0.2 | 330±23 | 41±7.5 | 0.49±0.13 | 574±275 |
| Pt | 1.81±0.1 | 227±50 | 25±0.9 | 0.10±0.03 | 215±115 |
| $MoO_x$ (thin) | 8.91±2.5 | 309±61 | 31±3.1 | 0.86±0.34 | 113±14 |
| $MoO_x$ (thick) | 3.51±0.6 | 312±80 | 36±7.0 | 0.4±0.17 | 550±234 |

The $CuSbS_2$ PV devices on W, Ni, Pd, Au and FTO back contacts did not show any photoresponse, but for different reasons. The growth on Au resulted in a strong chemical reaction with $CuSbS_2$. This is surprising given gold's tendency to resist chemical reactions, but considering gold and copper are both group-11 elements, perhaps gold can replace copper and then more easily react with the $CuSbS_2$ lattice. The morphology of $CuSbS_2$ grown on Au was showed micron-sized spikes (supplementary Fig. S2), accompanied by large increase in conductivity, resulting in linear/shunted JV response of the PV devices. In contrast, JV device measurements of the $CuSbS_2$ films on W indicated large bulk resistance of the absorber material and still no photoresponse. The growth on Ni resulted in delimitation of the $CuSbS_2$ film, but in a slightly different way than for Pd and Pt (high temperature). Rather than curling up in small 0.1 mm flakes that are indicative of stress (on Pt and Pd), the $CuSbS_2$ films on Ni delaminated as flat 2 – 4 mm flakes. Finally, the as-grown $CuSbS_2$ absorber films on fluorine doped tin oxide (FTO) had high morphological quality, but within hours pinholes started to form, growing in size and density and eventually resulting in the loss of the films.

In addition to metallic back contacts, we also studied $MoO_x$ and $Cu_{12}Sb_4S_{13}$ as charge-selective layers on Mo electrodes (Table II), similar to what has been previously studied for CdTe [37] and CIGS [38]. The PV devices grown on Mo/$MoO_x$ had better efficiency (up to 1%) due to higher $J_{sc}$ (up to 10 mA/cm$^2$), but only for thinner (0.8 $\mu$m) $CuSbS_2$ absorber layers (Table II). The $J_{sc}$ difference with the thicker (1.2 $\mu$m) $CuSbS_2$ absorber layers can be explained by the $MoO_x$ charge-selective properties (Fig. 8). $MoO_x$ has a deep (6.6eV) work function [39] that is suitable to collect holes better than Mo (Fig. 8a), and thus reflects photogenerated electrons due to the resulting upwards band-bending in the absorber. This turned out to be particularly important for $J_{sc}$ of the PV devices with thin $CuSbS_2$, where the electrons are generated throughout the thickness of the absorber (Fig. 8c). On the other hand, for the thick $CuSbS_2$ absorber, most of the electrons are generated close to the front contact, and hence $MoO_x$ does not affect the device performance (Fig.8b). Overall, these findings are similar to what was recently reported for backwall super-strate CIGS [40]. Similar $CuSbS_2$ charge-selective back contact experiments with 40nm of $Cu_{12}Sb_4S_{13}$ degenerate p-type semiconductor [15] did not lead to any statistically significant improvements in PV device performance, regardless of the thickness of the $CuSbS_2$ absorber (0.6-1.2$\mu$m), suggesting that the $Cu_{12}Sb_4S_{13}$ /$CuSbS_2$ valence band alignment may not be suitable for this device design.

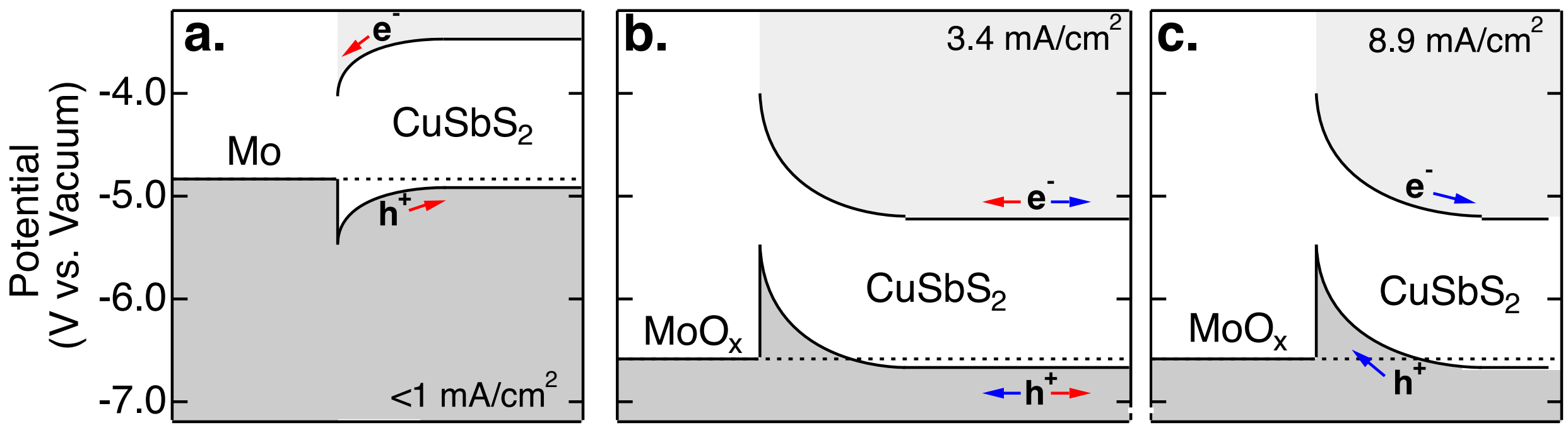

**Figure 8:** Schematic band diagram for (a) Mo/$CuSbS_2$ (thin), (b) $MoO_x$/$CuSbS_2$ (thick) and (c) $MoO_x$/$CuSbS_2$ (thin), showing enhanced photocurrent only for thinner $CuSbS_2$ absorber layers.

## 4. Summary and conclusions

Accelerated development of thin film photovoltaic device prototypes has been demonstrated on the example of novel $CuSbS_2$ absorbers, with substrate PV device architecture and a CdS heterojunction partner. First, reproducible $CuSbS_2$ synthesis at elevated temperatures has been achieved by introducing three-stage self-regulated absorber growth process, eliminating detrimental Cu-rich and Sb-rich competing impurity phases. Second, combinatorial PV device studies suggest that $CuSbS_2$ crystallographic orientation and the resulting morphology are important materials parameters that control the trade-off of the open circuit voltage and short circuit current of the solar cells. Third, the high-throughput experiments also indicate that the optimal thickness of the $CuSbS_2$ absorbers is 1.4 $\mu$m when deposited on Mo (better than Pt, Pd, Au, Ni, W, FTO), and 0.8 $\mu$m when deposited on $MoO_x$ charge-selective contact. Together, these three findings demonstrate the benefits of high-throughput combinatorial approach to accelerated development of initial PV device prototypes.

In conclusion, more research and development is needed to enhance the energy conversion efficiency of the $CuSbS_2$ PV technology beyond the ~1% demonstrated here. Currently, the device efficiency is limited by low short circuit current that results from poor collection of the photogenerated charge carriers. Identification and quantification of defects that limit minority carrier lifetimes would help to determine if the low electron diffusion lengths is due to the poor absorber morphology related to the deposition method, or due to intrinsic bulk defect properties of the $CuSbS_2$ material. Further research and development needs to include control of crystallographic orientation of the $CuSbS_2$ absorbers, and development of alternative $CuSbS_2$ heterojunction partners with higher conduction band position compared to CdS. Both of these efforts should result in improvements in open-circuit voltage and overall energy conversion efficiency of the $CuSbS_2$ PV technology.

## 5 Acknowledgments

The "Rapid Development of Earth-abundant Thin Film Solar Cells" project is supported by the U. S. Department of Energy, Office of Energy Efficiency and Renewable Energy, as a part of the SunShot initiative, under Contract No. DE-AC36-08GO28308 to NREL. LLB was supported by the Department of Defense through the National Defense Science and Engineering Graduate Fellowship. We would like to acknowledge our NREL colleagues, Ingrid Repins and Miguel Contreras for discussion and help with thin film chalcogenide device fabrication and characterization, Jeff Alleman, Steven Robbins, Danny Yerks for assistance minimizing chamber down time and building the JV mapping tool, and Bobby To for SEM characterization.

## Supplementary information

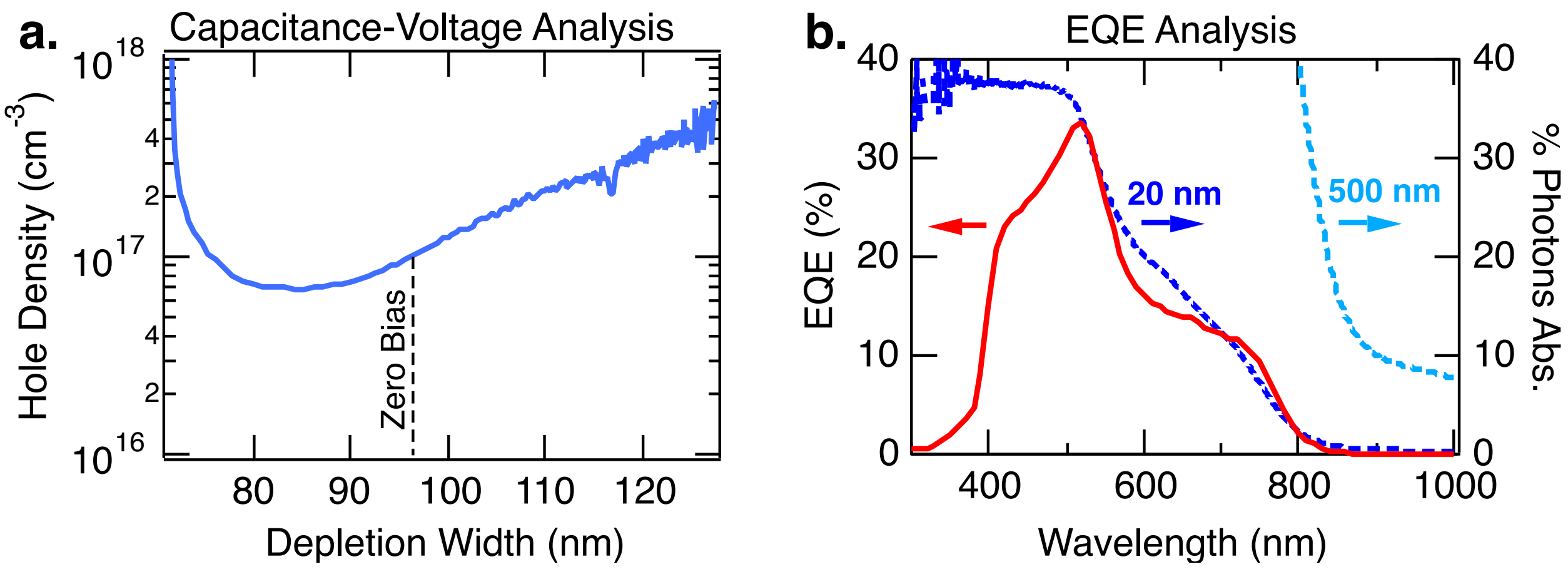

**Fig. S1** (a) The results of C-V measurements of the CuSbS2 PV devices, showing ~100 nm depletion width. (b) Modeled absorbance of CuSbS2 layers with 20 nm and 500 nm thickness, showing that the carriers are collected within 20 nm of the heterojunction contact.

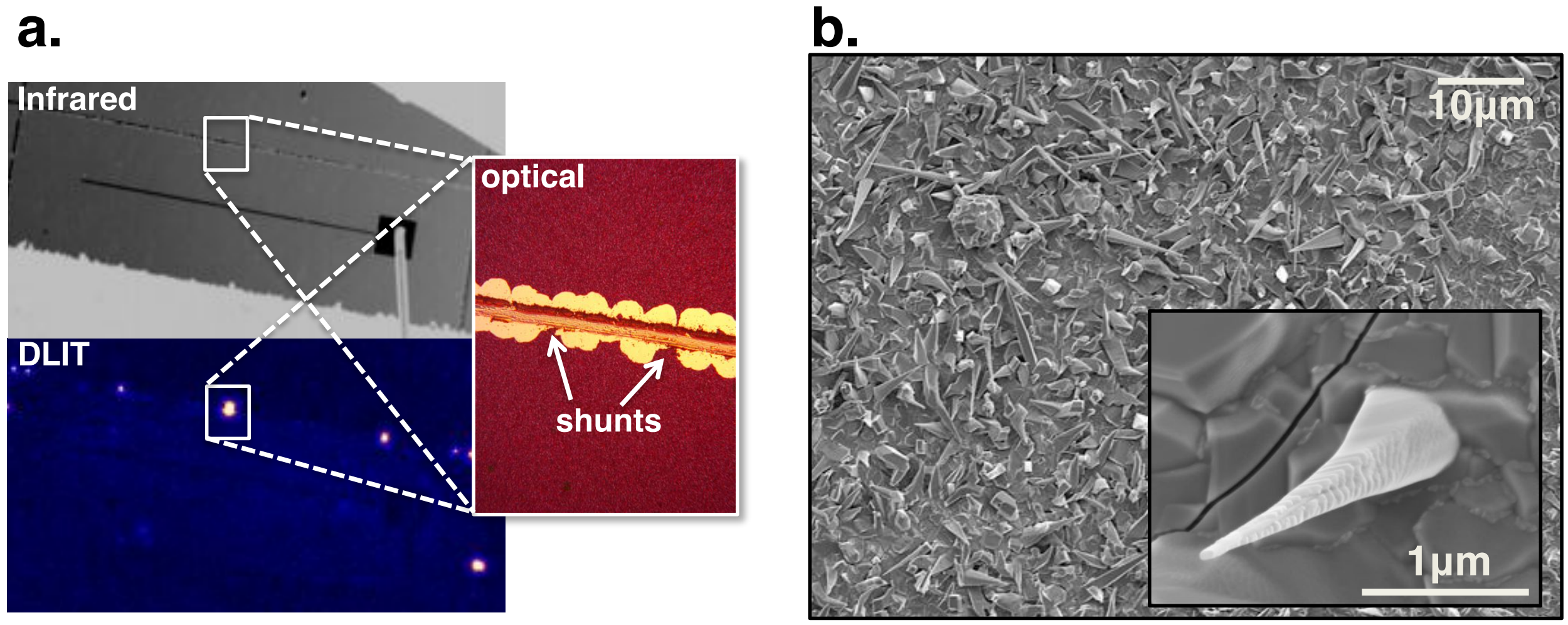

**Figure S2:** (a) Infrared, optical, and dark lock-in thermography images of shunting along the scribe edge of PV devices with Pt back contact. (b) Scanning electron microscopy images of micro-spike growth of CuSbS2 with Au back contact.